\pdfoutput=1
\documentclass[aps,prl,floatfix,twocolumn]{revtex4-1}  
\usepackage{physics}
\usepackage{amsfonts}
\usepackage{mathrsfs}
\usepackage{amsmath}
\usepackage{color}
\usepackage{graphicx}
\usepackage{bm}
\usepackage{amssymb}
\usepackage{xspace}
\usepackage{epstopdf}
\usepackage{dcolumn}
\usepackage{longtable}
\usepackage[colorlinks=true, letterpaper=true, pdfstartview=FitV, linkcolor=blue, citecolor=blue, urlcolor=blue]{hyperref}

\begin{document}
\title{Flavor Quantum Dots and Artificial Quark Model in Transition Metal Dichalcogenides}
\author{Zhi-qiang Bao}
\affiliation{Department of Physics, University of Texas at Dallas, Richardson, Texas 75080, USA}
\author{Patrick Cheung}
\affiliation{Department of Physics, University of Texas at Dallas, Richardson, Texas 75080, USA}
\author{Fan Zhang}\email{zhang@utdallas.edu}
\affiliation{Department of Physics, University of Texas at Dallas, Richardson, Texas 75080, USA}
\date{\today}
\begin{abstract}
We show that the triply degenerate $Q$ valleys in few-layer transition metal dichalcogenides 
provide a unique platform for exploring the rare flavor SU(3) symmetry in quantum dot geometry.
The single and double dots are reminiscent of the quark model and eightfold way, 
and their many-body triplets and octets may be regarded as artificial quarks and hadrons.
For the artificial quark transistor, each level hosts one central and two side Coulomb peaks 
of irrational height ratios, and flavor Kondo effects occur at $1/3$ and $2/3$ fillings 
with fractional conductance quantization in the unitary limit.
\end{abstract}
\maketitle

\indent\textcolor{blue}{\em Introduction.}---Progress in condensed matter physics is often 
driven by discovery of novel materials. 
In this regard, materials presenting unique symmetry and dimensionality are of particular importance. 
Atomically thin semiconducting transition metal dichalcogenides 
(TMDs)~\cite{Xu,Mak} represent a conceptually new class of 2D materials that have large 
band gaps, strong spin-orbit couplings (SOC), heavy effective masses, and multiple valley degrees of freedom. 
On this basis, TMDs provide a fertile ground for many-body physics in reduced dimensions 
that has never ceased to surprise us. 

\begin{figure}[!b]
\includegraphics[width=1.0\columnwidth]{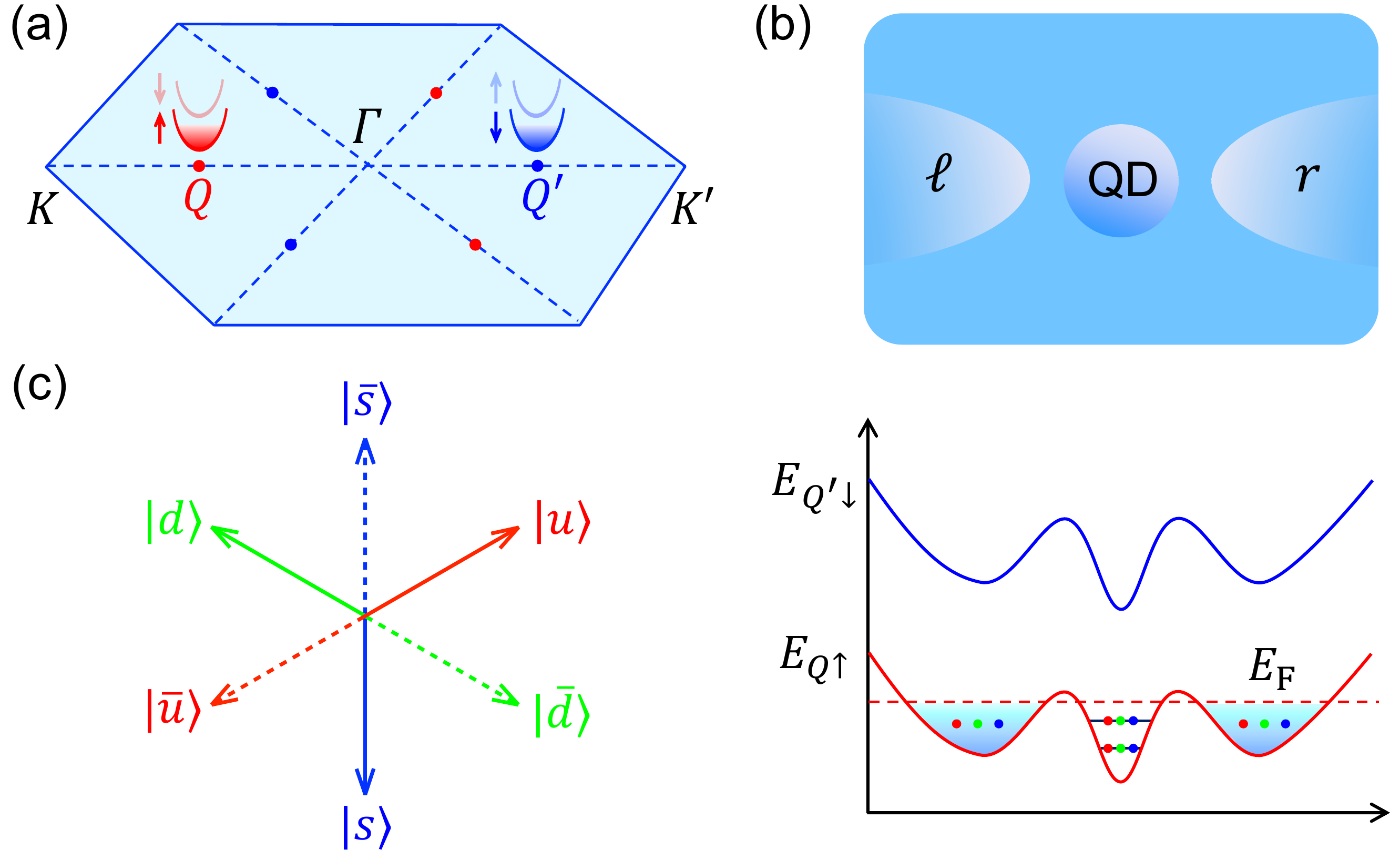}
\caption{(a) Hexagonal BZ of odd-layer TMD with spin-split conduction band. 
Three $Q$ valleys are related by $\mathcal{C}_{3}$ symmetry.
$Q$ and $Q'$ valleys are related by $\mathcal{T}$ symmetry.
(b) Gate defined QD device, with discrete energy levels,
tunnel-coupled to two electrodes.  
(c) Flavor symmetry of the three $Q$ valleys dubbed $u$, $d$, and $s$. 
For a singly ($1/3$) occupied level, the state can be $|u\rangle$, $|d\rangle$, $|s\rangle$, 
or their linear combination, forming a flavor triplet. For a doubly ($2/3$) occupied level, the state can be 
$|\bar{u}\rangle$, $|\bar{d}\rangle$, $|\bar{s}\rangle$, or their linear combination, forming an antitriplet. 
}
\label{fig1}
\end{figure}

Fig.~\ref{fig1}(a) depicts the hexagonal first Brillouin zone (BZ) of few-layer TMDs. 
There are two inequivalent hexagon corners, $K$/$K'$ points. 
By contrast, the six $Q$/$Q'$ points in between $\Gamma K$/$\Gamma K'$ are all inequivalent. 
The three $Q$ points are related by threefold rotational ($\mathcal{C}_{3}$) symmetry, 
and the $Q$ and $Q'$ points are related by time reversal ($\mathcal{T}$) symmetry. 
Uniquely, the $Q$/$Q'$ points are the conduction band minima of few-layer TMDs~\cite{Wu,Pisoni,Chen}. 
For odd-layers the bands in each $Q$ valley are spin split due to the large SOC 
and the intrinsic inversion ($\mathcal{P}$) asymmetry. In addition, 
because of mirror ($\mathcal{M}_{z}$) symmetry 
the $z$-component of spin is a good quantum number. 
Markedly, the $Q$ valley sub-bands exhibit both spin splitting and spin-valley locking.  
By contrast, for even-layers the restored $\mathcal{P}$ symmetry 
requires all states to be Kramers degenerate. 
Such universal even-odd layer-dependent band structures have been theoretically elucidated~\cite{Wu} 
and experimentally observed~\cite{Wu,Pisoni,Chen} in the Shubnikov-de Hass oscillations. 
At small fields, the oscillations display a $6$ ($12$) fold Landau-level degeneracy 
in the odd- (even-) layers because of 
$\mathcal{C}_{3}\times\mathcal{T}$ ($\mathcal{C}_{3}\times\mathcal{T}\times\mathcal{P}$) 
symmetries. At moderate fields, an inter- (intra-) valley Zeeman effect occurs in the odd- (even-) 
layers, reducing the degeneracy from $6$ ($12$) to $3$ ($6$). 

Therefore, the $n$-type few-layer TMDs offer a unique opportunity to create 
quantum dots (QDs) with an emergent flavor SU(3) symmetry and to explore how the flavor symmetry  
give rise to new transport phenomena under the influence of many-body interactions. 
Historically, a close examination of this symmetry led directly to the discovery of 
the quark model and eightfold way~\cite{Greiner}. 
However, the flavor SU(3) symmetry is rare in solid state systems, 
although key progresses have been made on the SU(4) valley QD and 
the SU(N) QD arrays~\cite{SU4,Oreg,crossover,Teratani}.
Recently, the gate-defined QD devices based on few-layer TMDs have been successfully  fabricated~\cite{Song,Guo,Wang,Ensslin,Epping}. 

Here we demonstrate in QD physics that the triply degenerate TMD $Q$ valleys 
provide an unprecedented platform for exploring the rare flavor SU(3) symmetry.
The construction of single and double QD states are reminiscent of the quark model and eightfold way, 
and the QD triplets and octets may be regarded as artificial quarks and hadrons, 
analogous to the concept of artificial atoms~\cite{Kouwenhoven4,Kouwenhoven3} in spin-QDs.
For an artificial quark transistor, each level hosts one central and two side Coulomb peaks 
of irrational height ratios, and two flavor Kondo effects occur at one-electron and one-hole fillings 
with fractional conductance quantization.
In sharp contrast to those of spin-QDs~\cite{Goldhaber-Gordon,Cronenwett,Kouwenhoven1,Kouwenhoven2},  
our results establish a framework for flavor QDs and artificial quark model, 
thereby paving the way for exploring the interplay of symmetry, geometry, and interaction in 2D materials.

\indent\textcolor{blue}{\em Device and model.}---In Fig.~1(b), we construct a QD device based on 
an $n$-type odd-layer TMD. The source and drain electrodes and the QD region can be defined by multiple gates laterally. 
(Vertical QD geometry is not excluded.)
The confining potential, tunable by the gates, leads to discrete levels in the QD and continuum states in the electrodes. 
The low-energy Hamiltonian of each $Q$ valley can be expressed as $\mathcal{H}=p_x^2/2m_x+p_y^2/2m_y$ 
with the anisotropic effective masses $m_{x,y}\sim 0.5m_e$~\cite{Wu,Q-QHF}. The confining potential may be modeled by 
$V=m_e\omega r^2/2$, which is sufficiently smooth to diminish the intervalley scattering 
and sufficiently isotropic to respect the rotational symmetry. This yields sixfold degenerate discrete QD levels 
$\epsilon_{n_x,n_y}=\hbar\omega\sqrt{m_e}[(n_x+1/2)/\sqrt{m_x}+(n_y+1/2)/\sqrt{m_y}]$, 
where $n_x$ and $n_y$ are non negative integers and typically $\hbar\omega\gtrsim 1$~meV~\cite{Kouwenhoven3}.
When a perpendicular magnetic field $\sim 5$~T is applied, an intervalley Zeeman effect~\cite{Wu} 
lifts the spin degeneracy yet preserves the rotational symmetry. 
The resulting Zeeman splitting is typically $\gtrsim 1$~meV in transport experiments~\cite{Wu,Pisoni,Chen}.

Therefore, each QD level has an emergent flavor SU(3) symmetry, arising from the rotational symmetry among 
the three $Q$ valleys of the same spin polarization, and the QD level spacing is $\Delta\gtrsim 1$~meV. 
(The same flavor symmetry also applies to each electrode continuum.) 
When electron-electron interactions are taken into account, 
the addition of each electron into the QD costs an extra Coulomb energy. 
Mesoscopically, this can be modeled by the single-electron charging energy 
$E_{C}=e^{2}/2C$, where $C$ is the effective QD capacitance. Since typically $E_C>\Delta$ 
in experiments~\cite{Kouwenhoven3,Guo,Ensslin}, $E_c$ is the dominant energy scale in QD physics.
Importantly, the number of electrons in the QD is tunable by a back gate. 

\indent\textcolor{blue}{\em Artificial quarks.}---Label the three symmetric $Q$ valleys by $u$, $d$, and $s$, 
and consider an isolated subspace formed by the possible states of the topmost QD level. 
For the case of one electron occupying the level, the state can be $|u\rangle$, $|d\rangle$, $|s\rangle$, 
or their linear combination, exhibiting a flavor SU(3) symmetry. 
For the case of two electrons occupying the level, the state can be $|ds\rangle$, $|su\rangle$, $|ud\rangle$, 
or their linear combination, also exhibiting the flavor SU(3) symmetry. Intriguingly, 
the two-electron states amount to the one-hole states, i.e., $|\bar{u}\rangle\equiv|ds\rangle$, 
$|\bar{d}\rangle\equiv|su\rangle$, and $|\bar{s}\rangle\equiv|ud\rangle$. Likewise, 
the zero-electron and three-electron states are $|0\rangle$ and $|\bar{0}\rangle\equiv|uds\rangle$, 
respectively, both of which are unique and flavorless. Naturally, there is a particle-hole symmetry in each subspace. 

Fig.~1(c) sketches these states and their flavor SU(3) symmetry, which is analogous to the quark model~\cite{Greiner}. 
The one-electron states, i.e., the artificial quarks $|u\rangle$, $|d\rangle$, and $|s\rangle$, 
constitute a triplet and form the $D^{10}$ irreducible representation of the flavor SU(3) group. 
By contrast, the two-electron (or one-hole) states, i.e., the artificial antiquarks 
$|\bar{u}\rangle$, $|\bar{d}\rangle$, and $|\bar{s}\rangle$, 
constitute an antitriplet and form the $D^{01}$ irreducible representation. 

\begin{figure}[!t]
\includegraphics[width=1.0\columnwidth]{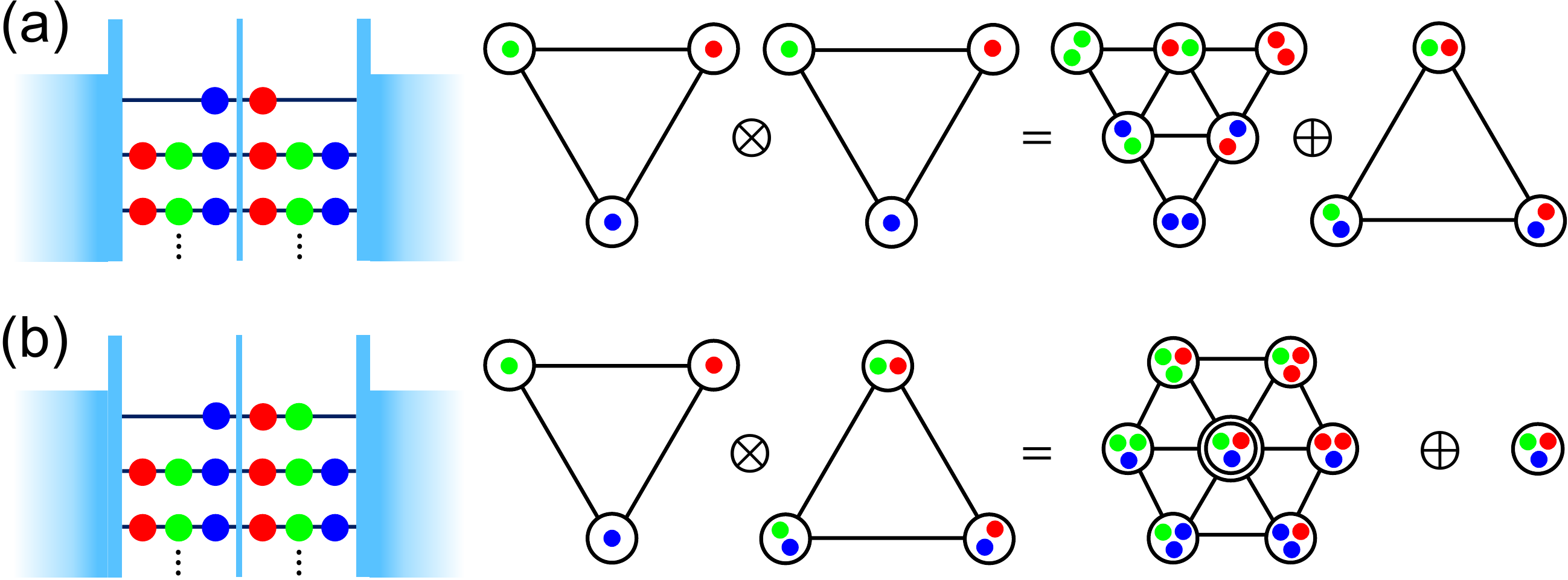}
\caption{(a) $(1,1)$ double flavor-QDs and its composite sextet and antitriplet. 
(b) $(1,2)$ double flavor-QDs and the composite octet and singlet. 
1 and 2 denote respectively the single and double occupation of the topmost level of a QD.}
\label{fig2}
\end{figure}

\indent\textcolor{blue}{\em Artificial hadrons.}---After establishing the many-body states 
of single flavor-QD, we further examine the composite many-body states of symmetric double flavor-QDs. 
Experimentally, double QDs have already been fabricated in few-layer MoS$_{2}$ devices~\cite{Guo,Ensslin}. 
For the SU(2) case, double spin-QDs can form spin singlet and triplet states, 
which follows from the tensor product decomposition rule: $2\otimes 2=3\oplus 1$,
where $1$, $2$, and $3$ denote respectively the spin singlet, doublet, and triplet. 
However, for the SU(3) multiplets, the tensor product decomposition rules are fundamentally different~\cite{Greiner}, e.g.,  
\begin{equation}\label{add}
3\otimes 3=6\oplus\bar{3}\,,\quad 3\otimes\bar{3}=8\oplus 1\,.
\end{equation}
The $3$ and $\bar 3$ represent the flavor triplet and antitriplet, respectively,    
which can be formed by the light quarks ($u$, $d$, and $s$) and antiquarks ($\bar u$, $\bar d$, and $\bar s$),
The $1$, $6$, and $8$ represent the flavor singlet, sextet, and octet, respectively, 
forming the $D^{00}$, $D^{20}$, and $D^{11}$ irreducible representations. 
Historically, the $8$ predicted the eightfold way of baryon and meson octets~\cite{Greiner}. 
These fundamentals already imply the novelty, significance, and richness of flavor-QD physics based on TMD $Q$ valleys.
   
Fig.~\ref{fig2}(a) sketches the $(1,1)$ double QDs in which the topmost level of each is singly occupied.
In light of Eq. (\ref{add}), the composite states are the antitriplet and sextet,
in sharp contrast to the singlet and triplet in the spin case. 
Fig.~\ref{fig2}(b) sketches the $(1,2)$ double QDs 
in which their topmost levels are singly and doubly occupied, respectively. 
Based on Eq.~(\ref{add}), the composite states are the singlet and octet. 
This octet of three artificial quarks is analogous to the baryon octet.
As the doubly occupied level is an antitriplet,  
this octet (of one artificial quark and one artificial antiquark) is also analogous to the meson octet.
Similarly, we can consider the $(2,1)$ and $(2,2)$ double QDs, which yield the antiparticles of 
the multiplets in Fig.~\ref{fig2} because of the particle-hole symmetry.

\begin{figure}[!t]
\includegraphics[width=1.0\columnwidth]{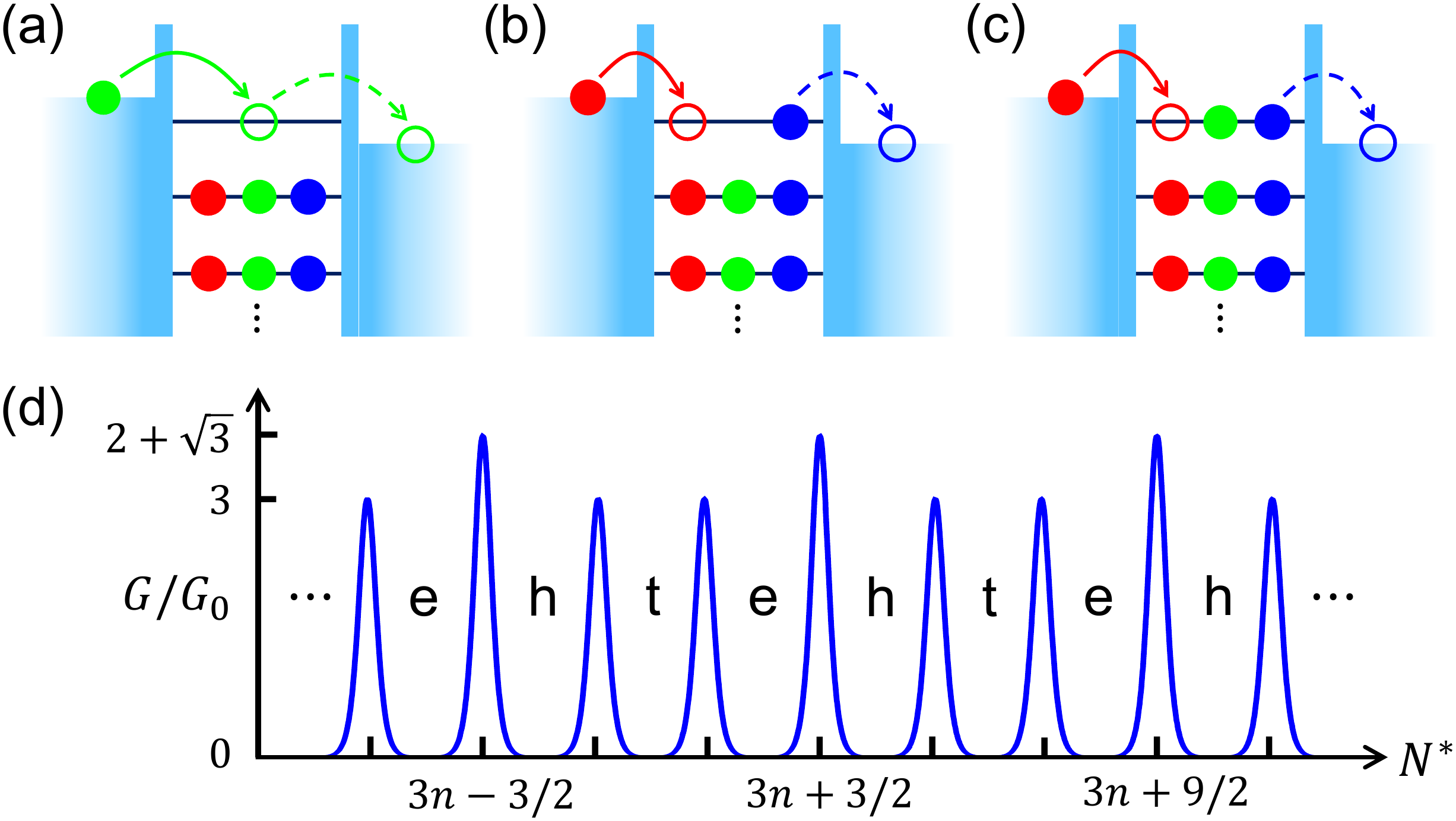}
\caption{(a)-(c) Sequential tunnelings in a flavor-QD under a small source-drain bias, 
leading to a conductance valley (peak) in (d) when the QD occupation number $N^*$ is an integer (half integer).
(d) Differential conductance $G$ versus $N^*$ at temperature $k_{B}T=0.1E_{C}$. 
$e$, $h$, and $t$ respectively denote the QD states with $3n\pm 1$ and $3n$ confined electrons.}
\label{fig3}
\end{figure}

\indent\textcolor{blue}{\em Coulomb blockade.}---The artificial quarks and hadrons can be probed by transport 
when weak tunnel couplings are allowed between the flavor-QD(s) and nearby source and drain electrodes, 
which can give rise to novel transport anomalies such as flavor enriched Coulomb blockades and flavor enforced Kondo effects. 
Here we focus on examining the Coulomb peaks and Kondo effects in the single QD device 
and elucidating the roles of flavor SU(3) symmetry in their smoking-gun signatures. 

The QD is transparent if the source-drain bias voltage $eV_{SD}\geq E_{C}$, 
and the conductance $G$ is thermally diminished if $k_{B}T\geq eV_{SD}$. 
Thus, we focus on the regime in which $k_{B}T\ll eV_{SD}\ll E_{C}$. 
This implies that either no or one QD level lies between the source and drain potentials. 
In the former case, $G$ vanishes since no level mediates transport,
i.e., Coulomb blockade~\cite{Beenakker,Glazman}. In the latter case, 
$G$ is peaked, and Figs.~\ref{fig3}(a)-\ref{fig3}(c) depict its three scenarios. 
It is enlightening to interpret the left-to-right electron flow in Fig.~\ref{fig3}(a) 
and the right-to-left hole flow in Fig.~\ref{fig3}(c) as the transport of artificial quarks 
and antiquarks. Intriguingly, the charge flow in Fig.~\ref{fig3}(b) is  
attributed to the artificial quark-antiquark resonance of the QD state.
Such an analogy suggests that the conductances in Fig.~\ref{fig3}(a) and Fig.~\ref{fig3}(c) 
are the same because of the particle-hole symmetry, 
and that they are lower than the resonance conductance in Fig.~\ref{fig3}(b). 

Such an artificial quark transistor can be described by 
\begin{align}\label{Ht}
\mathcal{H}=&\sum_{\bm{k}f\alpha}\varepsilon_{\bm{k}\alpha}c_{\bm{k}f\alpha}^{\dagger}c_{\bm{k}f\alpha}
+\sum_{\bm{k}f\alpha}\left(t_{\alpha}c_{\bm{k}f\alpha}^{\dagger}d_{mf}+\mbox{H.c.}\right)\nonumber\\
&+\sum_{mf}\epsilon_m d_{mf}^{\dagger}d_{mf} 
+E_C\left({N}-N^*\right)^2.
\end{align}
$c_{\mathbf{k}f\alpha}$ is the annihilation operator of an electron of energy $\varepsilon_{\mathbf{k}\alpha}$ 
and flavor $f$ ($u, d, s$) in the lead $\alpha$ ($\ell, r$). 
$d_{mf}$ is the annihilation operator of an electron of energy $\epsilon$ and flavor $f$ in the QD level $m$.
$N$ is the total number of electrons in the QD, and $N^*$ is tunable via a back gate.
The tunneling matrix elements $t_{\alpha}$ are real, 
flavor-momentum independent, and much smaller than $E_C$.
Further, we denote by $P_f$ the probability of the QD level to be occupied with one electron of flavor $f$, 
by $P_{\bar f}$ to be occupied with two electrons of flavors not being $f$ (i.e. one hole of flavor $f$), 
by $P_0$ to be empty, and by $P_{\bar 0}$ to be occupied with three electrons. 
In the Fermi Golden rule approximation, the dynamic behaviors of these probabilities can be 
determined by rate equations~\cite{Beenakker,Glazman}. 

For $N^*= 3n+1/2$ in Fig.~\ref{fig3}(a), the state with $3n$ electrons and 
the three states with $3n+1$ electrons are in resonance, 
and their probabilities obey the rate equations
\begin{equation}\label{CP1}
\dot P_0=\sum_{f}\left(\Gamma_{o}P_f-\Gamma_{i}P_0\right),\quad
\dot P_f=\Gamma_{i}P_0-\Gamma_{o}P_f,
\end{equation}
and the condition $P_0+\sum_{f}P_f=1$.
Here $\Gamma_{i,o}=\Gamma_{i,o}^{\ell}+\Gamma_{i,o}^{r}$ are the transition rates due to 
the tunneling of one electron into ($i$) and off ($o$) the QD, 
with $\Gamma_{i}^{\alpha}=\Gamma^{\alpha}n_{\rm F}^{\alpha}/\hbar$, 
$\Gamma_{o}^{\alpha}=\Gamma^{\alpha}(1-n_{\rm F}^{\alpha})/\hbar$, 
and $\Gamma^{\alpha}=2\pi|t_{\alpha}|^2\nu_{\alpha}$.
The current through the QD can be found as the one through the left junction:  
$I=-e\sum_{f}\left(\Gamma^{\ell}_{i}P_0-\Gamma^{\ell}_{o}P_f\right)$.
In the steady state, $\dot P_0=\dot P_f=0$, and the conductance $dI/dV$ reads  
\begin{equation}\label{G1}
\!\!G_{3n+\frac{1}{2}}(x)=\frac{e^2}{\hbar}\frac{\Gamma^{\ell}\Gamma^{r}}{\Gamma^{\ell}+\Gamma^{r}}
\frac{1}{k_{\rm B}T}\left[\frac{-3{n_{\rm F}}'(x)}{1+2n_{\rm F}(x)}\right],
\end{equation}
where $n_{\rm F}(x)=1/(e^x+1)$ and $x=2E_C(N-N^*)/k_{\rm B}T$. 
We note that $G_{3n+\frac{5}{2}}(x)=G_{3n+\frac{1}{2}}(-x)$ because of the particle-hole symmetry.
For $N^*= 3n+3/2$ in Fig.~\ref{fig3}(b), the six states with $3n+1$ or $3n+2$ electrons are in resonance, 
and their probabilities are governed by the condition 
$\sum_{f}\left(P_f+P_{\bar f}\right)=1$ and the rate equations 
\begin{equation}\label{CP2}
\!\!\dot P_f=\!\!\sum_{f'}\!'\!\left(\Gamma_{o}P_{\bar{f'}}-\Gamma_{i}P_f\right),\;
\dot P_{\bar f}=\!\!\sum_{f'}\!{'}\!\left(\Gamma_{i}P_{f'}-\Gamma_{o}P_{\bar{f}}\right),
\end{equation}
where $\sum'$ means $f'\neq f$. The current through the QD is  
$I=-e\sum'_{f'}\sum_{f}\left(\Gamma^{\ell}_{i}P_f-\Gamma^{\ell}_{o}P_{\bar f}\right)$.
In the steady state, $\dot P_f=\dot P_{\bar f}=0$, we obtain the conductance $dI/dV$
\begin{equation}\label{G2}
G_{3n+\frac{3}{2}}(x)=\frac{e^2}{\hbar}\frac{\Gamma^{\ell}\Gamma^{r}}
{\Gamma^{\ell}+\Gamma^{r}}\frac{1}{k_{\rm B}T}\left[-2{n_{\rm F}}'(x)\right].
\end{equation}

The bracket in Eq.~(\ref{G2}) reaches its maximal value $1/2$ at $x=0$, 
whereas the bracket in Eq.~(\ref{G1}) reaches its maximal value $3/(\sqrt{3}+1)^2$ at $x=\ln\sqrt{3}$. 
Therefore, the three Coulomb peaks of a given flavor-QD level 
have two important characteristics, as shown in Fig.~\ref{fig3}(d).
First, their heights have irrational ratios
\begin{equation}\label{ratio}
G_{3n+\frac{1}{2}}^{\rm max}:G_{3n+\frac{3}{2}}^{\rm max}:G_{3n+\frac{5}{2}}^{\rm max}
=3:(2+\sqrt{3}):3\,.
\end{equation}
Second, while the central peak is symmetric around $N^*=3n+3/2$, the two side peaks 
are shifted outward by $(k_{\rm B}T/4E_C)\ln 3\,$ from $\,N^*=3n+1/2$ and $3n+5/2$, respectively. 
Although the side peaks are asymmetric around their maxima, the three-peak pattern is symmetric, 
as dictated by the particle-hole symmetry.

\begin{figure}[!t]
\includegraphics[width=1.0\columnwidth]{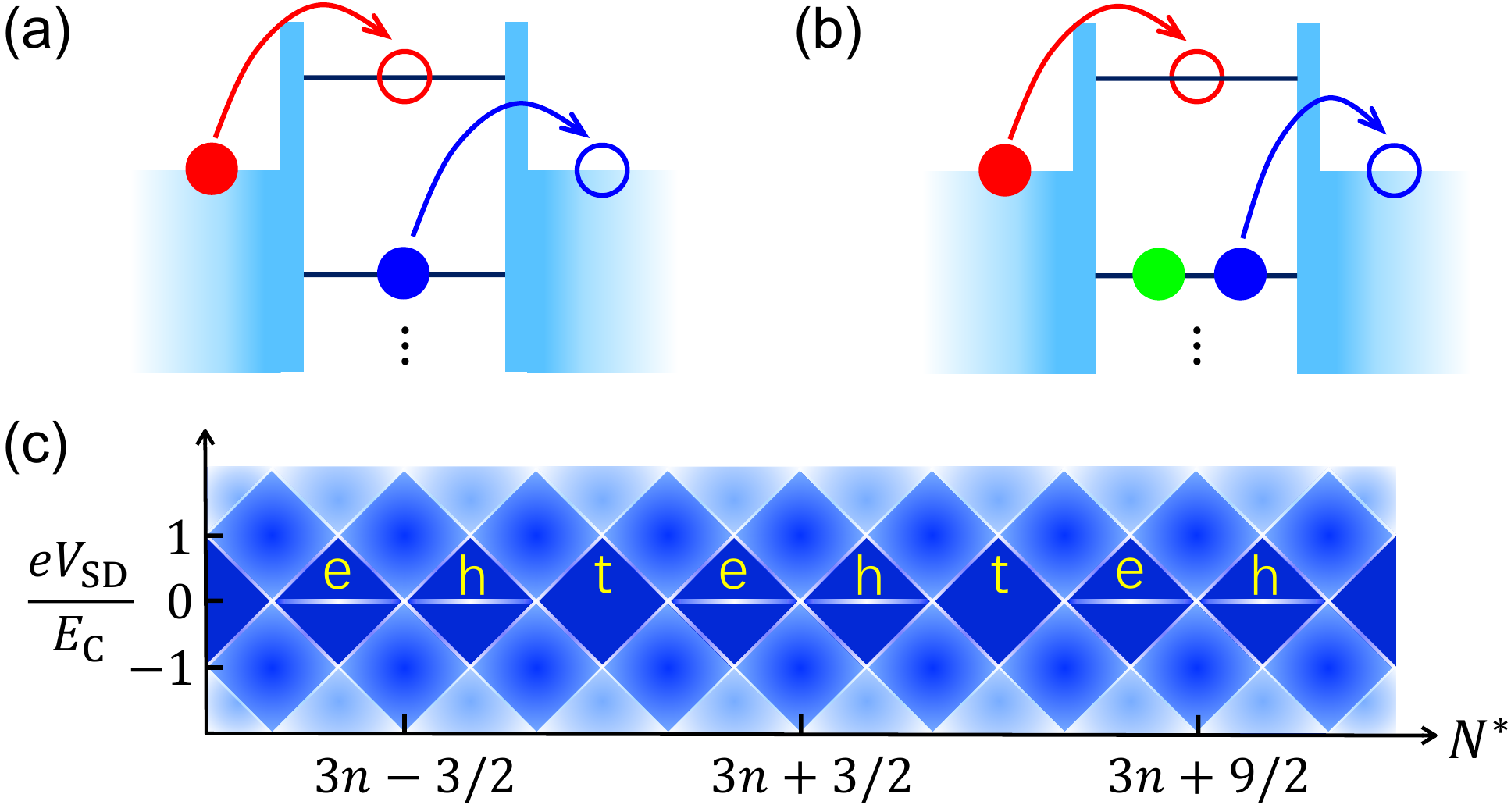}
\caption{Examples of co-tunneling events changing the flavors of QD states, 
for the topmost QD level being (a) singly ($1/3$) occupied or (b) doubly ($2/3$) occupied.
(c) Schematic $dI/dV$ versus $N^*$ (tunable by a back gate) and bias voltage $V_{SD}$ for a flavor-QD. 
White (blue) color denotes high (zero) conductance. Near zero bias, 
Coulomb peaks develop when $N^*$ is a half integer, Coulomb blockade exists when $N^*=3n$ (t), 
and Kondo effects occurs when $N^*=3n\pm 1$ (e and h).}
\label{fig4}
\end{figure}

\indent\textcolor{blue}{\em Kondo effects.}---Now we examine the flavor Kondo effects when the topmost 
QD level is singly ($1/3$) or doubly ($2/3$) occupied, as depicted in Figs.~\ref{fig4}(a)-\ref{fig4}(b).
Since the two cases are related by the particle-hole symmetry, we focus on the former. 
First, we perform the Schrieffer-Wolff transformation~\cite{Schrieffer}  
to project Eq.~(\ref{Ht}) into the concerned subspace up to second order in $t$. This 
gives rise to the flavor exchange model 
\begin{equation}\label{Hsd}
\mathcal{H}=\sum_{\bm{k}f}\varepsilon_{\bm{k}}c_{\bm{k}f}^{\dagger}c_{\bm{k}f}
+\sum_{\nu}J_{\nu\nu}f^{\nu}F^{\nu}\,,
\end{equation}
where $J_{\nu\nu}=t^2/E_C$, $t^2=t_{\ell}^2+t_r^2$, 
$c_{\bm{k}f}=(t_{\ell}c_{\bm{k}f\ell}+t_r c_{{\bm{k}} f r})/t$.
${\bm F}=\sum_{ff'}d^{\dag}_{f}{\bm\Lambda}_{ff'}d_{f'}$ and 
${\bm f}=\sum_{ff'}\sum_{{\bm k}{\bm k}'}c^{\dag}_{{\bm k}f}{\bm\lambda}_{ff'}c_{{\bm k}'f'}$ 
are the flavors of the QD electron and the lead electrons, respectively, 
with $\bm\lambda$ ($\bm\Lambda$) the standard eight Gell-Mann matrices~\cite{Greiner} 
for lead (QD) electrons.

Next, we compute the standard one-loop (1L) Feynman diagrams~\cite{Anderson,Zawadowski,Lindgren1,Lindgren2,Kuramoto1,Kuramoto2} 
to analyze the flows of exchange couplings $J_{\mu\nu}$. 
Using the renormalization group, we obtain the following set of flow equations
\begin{align}\label{flows}
\frac{d h^{(1L)}}{d\ell}=-\frac{1}{2}\sum_{\mu\nu}\sum_{\mu'\nu'}J_{\mu\nu}J_{\mu'\nu'}
[\lambda^{\mu},\lambda^{\mu'}][\Lambda^{\nu},\Lambda^{\nu'}]\,,
\end{align}
which leads to a nontrivial fixed point in which $J_{\mu\nu}=J\delta_{\mu\nu}$ and $J\rightarrow\infty$, 
in addition to the trivial one ($J_{\mu\nu}=0)$. 
It follows that the Kondo ground state must be a flavor singlet, 
similar to Nozi\`{e}res and Blandin's result on Kondo effects in metals~\cite{Blandin}.
Given the tensor product decomposition rules in Eq.~(\ref{add}), 
the singlet may be obtained by $\bar{3}\otimes{3}$ or $({3}\otimes{3})\otimes{3}$.
Physically, this means that the QD flavor of one artificial quark must be neutralized by one artificial antiquark (one hole) 
or two artificial quarks (two electrons) from the leads.
This suggests that the electron Kondo effect in Fig.~\ref{fig4}(a) can be viewed as 
QD flavor neutralization due to the right-to-left hole flow. 
(The flavor of lead electrons $c^{\dag}_{{\bm k}}{\bm\lambda}c_{{\bm k}'}$ 
can be expressed as $-c_{{\bm k}'}{\bm\lambda}^*c_{{\bm k}}^\dag$, 
where ${-\bm\lambda}^*$ is the conjugate representation of the Gell-Mann matrices~\cite{Teratani}.)

Therefore, the Friedel sum rule~\cite{Friedel,Nozieres} must dictate the total phase shift
to be $\sum_{f}\delta_{f}=-\pi$ or $2\pi$.
Given the flavor symmetry, we can conclude $\delta_{f}=-\pi/3$ or $2\pi/3$.
Based on the standard Landauer-B\"{u}ttiker formalism~\cite{Meir,Glazman}, 
the Kondo conductance can be expressed as 
\begin{align}\label{GKondo}
G=\frac{e^{2}}{h}\frac{4t_{\ell}^2t_{r}^2}{(t_{\ell}^{2}+t_{r}^{2})^2}
\sum_{f}\sin^{2}\delta_{f}\,.
\end{align}
Both $\delta_{f}=-\pi/3$ and $2\pi/3$ yield ${G}=9e^{2}/4h$ in the unitary limit ($t_{\ell}=t_r$),
validating the aforementioned two pictures for the QD flavor neutralization in the Kondo effect. 
Remarkably, $G$ can be larger than $2e^{2}/h$, the unitary spin Kondo conductance. 

For the hole Kondo effect in Fig.~\ref{fig4}(b), because of the particle-hole symmetry, 
it can be naturally viewed as QD flavor neutralization due to the left-to-right electron flow. 
Physically, this means that the QD flavor of one artificial antiquark can be neutralized 
by one artificial quark (one electron) from the leads. 
Thus, the phase shifts are $-\delta_{f}$, and the Kondo conductance is also given by Eq.~(\ref{GKondo}). 
Fig.~\ref{fig4}(c) features the e-h-t characteristic in the $dI/dV$ diamond structure for the flavor-QD, 
i.e., the electron and hole Kondo effects arise in two of the three Coulomb valleys per QD level, 
in sharp contrast to the even-odd characteristic of the spin-QD~\cite{Goldhaber-Gordon,Cronenwett,Kouwenhoven1,Kouwenhoven2}. 
Finally, as the QD flavor is neutralized below the Kondo temperature, 
the finite-temperature scaling of $G$ should obey the standard $T^2$ law of Fermi liquids.

\indent\textcolor{blue}{\em Discussions.}---A few comments are in order. 
Although there is a close analogy between the construction of single and double flavor-QD states 
and the quark model and eightfold way, there is a key difference: 
the strong interactions of quarks are described by a SU(3) gauge theory,  
yet the Coulomb interactions of electrons are described by a U(1) gauge theory. 
It would be intriguing to explore how to obtain an emergent SU(3) gauge symmetry 
in the proposed artificial quark model. For our purpose, the spin, valley, and layer SU(2) symmetries 
in TMD few-layers have been lifted. 
Richer flavor-QD physics is anticipated when these symmetries are retained.

Finally, $n$-type few-layer TMD is not the only material platform for the proposed physics. 
Both $p$-type few-layer hexagonal monochalcogenides MX (M=Ga, In; X=S, Se)~\cite{MX} 
and $(111)$ thin films of IV-VI semiconductors such as SnTe~\cite{SnTe} are candidate systems.
Coupled spin-QDs are the building blocks for semiconductor-based spin qubits and quantum computing~\cite{Wiel,Shulman}.
The proposed flavor-QDs may provide a unique way to study 2D materials-based qutrit (quantum trit) 
and open a new avenue for quantum information and computing.

\indent{\color{blue}\em Acknowledgment.}---We thank Marc Bockrath, Philip Kim, Kin Fai Mak, 
and Joe Qiu for valuable discussions. This work is supported by ARO under grant number W911NF-18-1-0416. 

\end{document}